\documentclass[useAMS,usenatbib]{mn2e}

\usepackage{graphicx}
\usepackage{amsmath}
\usepackage{amssymb}
\usepackage{color}
\usepackage{subfig}

\usepackage{hyperref}
\hypersetup{colorlinks=true,linkcolor=blue,citecolor=blue,filecolor=blue,urlcolor=blue}

\newcommand\msun{\, \rm M_\odot}

\newcommand\pc{{\, \rm pc}}

\newcommand\au{{\, \rm AU}}

\newcommand\aout{{a_{\rm out}}}

\title[IMRIs in Galactic Nuclei]{Intermediate-Mass Ratio Inspirals in Galactic Nuclei}
\author[G. Fragione and N. W. C. Leigh]{Giacomo Fragione$^{1}$\thanks{E-mail: giacomo.fragione@mail.huji.ac.il}, Nathan Leigh$^{2,3}$\\
$^1$Racah Institute for Physics, The Hebrew University, Jerusalem 91904, Israel\\
$^2$Department of Astrophysics, American Museum of Natural History, New York, NY 10024, USA\\
$^3$Department of Physics and Astronomy, Stony Brook University, Stony Brook, NY 11794-3800, USA}

\begin{document}

\maketitle

\begin{abstract}
In this paper, we study the secular dynamical evolution of binaries composed of intermediate-mass and stellar-mass black holes (IMBHs and SBHs, respectively) in orbit about a central super-massive black hole (SMBH) in galactic nuclei.  Such BH triplets could form via the inspiral of globular clusters toward galactic nuclei due to dynamical friction, or even major/minor galaxy mergers.  We perform, for reasonable initial conditions that we justify, sophisticated $N$-body simulations that include both regularization and Post-Newtonian corrections.  We find that mass segregation combined with Kozai-Lidov oscillations induced by the primary SMBH can effectively merge IMBH-SBH binaries on time-scales much shorter than gravitational wave emission alone.  Moreover, the rate of such extreme mass ratio inspirals could be high ($\sim 1\ \mathrm{Gpc}^{-3}\ \mathrm{yr}^{-1}$) in the local Universe, but these are expected to be associated with recent GC infall or major/minor mergers, making the observational signatures of such events (e.g., tidal debris) good diagnostics for searching for SMBH-IMBH-SBH mergers.  A small fraction could also be associated with tidal disruption events by the IMBH-SBH during inspiral.
\end{abstract}

\begin{keywords}
Galaxy: centre -- Galaxy: kinematics and dynamics -- stars: black holes -- stars: kinematics and dynamics -- galaxies: star clusters: general
\end{keywords}

\section{Introduction}

The long-standing question as to whether intermediate-mass black holes (IMBHs) ($100\ \mathrm{M}_{\odot}\lesssim M_{IMBH}\lesssim 10^5\ \mathrm{M}_{\odot}$) exist has recently come under intense scrutiny \citep{cann18,chili18,trem18,wrobel18}. While overwhelming evidence from both electromagnetic light and gravitational waves (GWs) has established the existence of Super Massive Black Holes (SMBHs; $M_{SMBH}\gtrsim 10^6\ \mathrm{M}_{\odot}$) in galactic nuclei \citep{kor13} and stellar-mass black holes (SBHs, $10\ \mathrm{M}_{\odot}\lesssim M\lesssim 100\ \mathrm{M}_{\odot}$) \citep{abb16}, the existence of IMBHs at any redshift is still highly debated \citep{mez17}.

Several frameworks have been suggested for the formation of IMBHs. In dense star clusters, runaway stellar collisions in the core may give birth to a very-massive star with a mass up to a few percent of the total mass of the cluster, that later collapses to form an IMBH \citep{por00}. The characteristic time-scale for this process to occur depends sensitively on the initial concentration of the cluster \citep{gie15}. IMBHs may also form from the direct collapse of massive primordial Pop III stars \citep{mad01,wha12,woo17}, or direct accretion on to SBHs \citep{leigh13,giersz15}.

However they form, if IMBHs exist they could be present in galactic nuclei. If the nucleus hosts an SMBH, a binary SMBH-IMBH system would then likely form \citep{pet17}. The delivery of IMBHs to galactic nuclei could be mediated by galaxy-galaxy mergers, gas accretion on to stellar-mass BHs in active galactic nuclei disks \citep{secunda18}, or inspiralling star clusters \citep*{mast14,fragk18,fragleiginkoc18}.

In our Galaxy, IMBHs have been claimed to be present in two globular clusters (GCs), i.e. $47$ Tuc \citep{kiz17} and $\omega$ Cen \citep{bau17}, on the basis of dynamical measurements. For IMBHs in extra-galactic clusters, the only way to observe them is if they happen to be accreting gas and producing associated high-energy photons. A few bright ultra-luminous X-ray sources ($10^{39}\ \mathrm{erg\ s}^{-1}\lesssim L_X\lesssim 10^{41}\ \mathrm{erg\ s}^{-1}$) can probably be explained by an accreting IMBH \citep{kaa17}. The recently observed tidal disruption event in an off-centre star cluster ($\sim 12.5$ kpc from the centre of the host galaxy) by \citet{lin18} is consistent with having been produced by an IMBH of mass $\sim 5\times 10^4\msun$.

Another distinctive signal of the presence of an IMBH could be GWs emitted if an SBH binary companion is inspiraling on to it \citep*{fragk18,fragleiginkoc18}. GW astronomy will therefore help significantly in the hunt for IMBHs. Present and upcoming GW facilities, such as LIGO\footnote{\url{http://www.ligo.org}}, \textit{LISA}\footnote{\url{https://lisa.nasa.gov}} and the Einstein Telescope\footnote{\url{http://www.et-gw.eu}} (ET) will be able to detect IMBH-SBH binaries of different masses. If GCs that harbour IMBH-SBH binaries are disrupted, e.g. due to galactic tides, any IMBH-SBH binaries will end up isolated in the field. However, some of these binaries may be delivered to the host galaxy nucleus before the host cluster disruption \citep{fragk18}, where they will interact with the local environment.  

In this paper, we study how IMBH-SBH binaries merge in galactic nuclei. We focus our attention on the IMRI rate due to the mergers of IMBH-SBH binaries driven by perturbations from the more massive SMBH, after addressing the expected details of the orbits characteristic of such binaries. We quantify the rates of such IMRI events by considering different SMBH-IMBH mass ratios and orbital parameters. We use high-precision direct $N$-body simulations, including Post-Newtonian terms up to PN2.5 order, to study the effects of the gravitational perturbations of the SMBH on the IMBH-SBH binary. We show that the strong tidal field of the primary SMBH may lead to high variations in the eccentricities and inclinations of the IMBH-SBH binaries, which may result in IMRI events.

The paper is organized as follows. In Sect. \ref{sect:imbhnuclei} we describe how IMBHs can be delivered to galactic nuclei. In Sect. \ref{sect:method}, we describe our numerical method to study IMBH-SBH mergers in galactic nuclei, while in Sect. \ref{sect:imris}, we describe our results. Finally, in Sect. \ref{sect:conc} we draw our conclusions.

\section{Delivering intermediate-mass black holes to galactic nuclei}
\label{sect:imbhnuclei}

Several mechanisms exist that could ultimately create BH triplets in galactic nuclei, both at home and abroad.  Below, we describe some of these mechanisms in more detail, and comment on the expected properties of any resulting BH triplets.

First, dynamical friction acting on GCs as they orbit through their host galaxies is thought to be able to deliver GCs to the centres of galaxies on time-scales much less than a Hubble time \citep[e.g.,][]{tremaine75,gne14}. If these GCs also host central IMBHs, then this could efficiently deliver these IMBHs to the outskirts of galactic nuclei \citep*{mast14,agu18,fragk18,fragleiginkoc18}. \citet{lei14} showed that, if stellar-mass BHs are also present in GCs hosting an IMBH, then at least one such SBH should usually be orbiting close to the IMBH on a bound roughly Keplerian orbit. By extrapolation, this predicts triplets of BHs in galactic nuclei that recently experienced the accretion of a GC hosting an IMBH.  These BH triplets should ultimately be formed out of the central SMBH and the inspiraling IMBH-SBH binary originating from its disintegrated GC host.  The details of the evolution of the SMBH-IMBH-SBH triplet depend on several competing effects, including mass segregation, secular dynamical effects induced by the central SMBH, direct interactions with single and binary stars, etc.  However, we emphasize that this formation mechanism for BH triplets predicts very small mass ratios.

Second, BH triplets could form due to major/minor mergers of galaxies \citep*[e.g.,][]{volon03}. Assuming that each progenitor galaxy hosts an SMBH-SMBH/SMBH-IMBH or IMBH-BH binary at its centre, then the mergers of galaxies could deliver two massive BH binaries in to close proximity, such that they might undergo a direct strong interaction.  Such four-body encounters could ultimately produce massive BH triplets, since a non-negligible fraction of chaotic four-body interactions are known to produce triples \citep[e.g.][]{leigh16} when typically the least massive object is ejected. This scenario ultimately predicts more massive BH triplets with mass ratios likely closer to unity, relative to the scenario described above.  

Third, BH triplets could form efficiently in the gaseous disks of active galactic nuclei (AGN) \citep[e.g.][]{secunda18}. In this scenario, a gaseous disk peppered with (initially) stellar-mass BHs orbits a central SMBH. If migration traps are present, differential gas torques exerted on the orbiting BHs will cause them to migrate towards a migration trap.  Once the first BH arrives in the migration trap, it sits there orbiting happily for an extended period of time, and in so doing stabilizes the orbits of other BHs migrating in toward it. This is accomplished via orbital resonances: migrating BHs that near the migration trap after the first BH end up locked in very high-order resonances, halting their migration. Turbulence in the gaseous disk can knock orbiting BHs out of resonance, however, allowing them to drift close to the trap and experience a close interaction with the first BH still orbiting there. The interaction is dissipative due to the gas, and it is possible that a SBH-SBH binary forms. Together with the central SMBH, this SBH-SBH binary forms a hierarchical BH triplet. The masses of the BHs are poorly constrained in this scenario, but an IMBH primary is possible since gas accretion within the disk can increase the stellar-mass BHs' masses considerably in some cases \citep{McKernan+2014}. Finally, this mechanism for forming BH triplets predicts roughly co-planar triplets, that should be Kozai-Lidov inactive.   

\begin{figure} 
\centering
\includegraphics[scale=0.4]{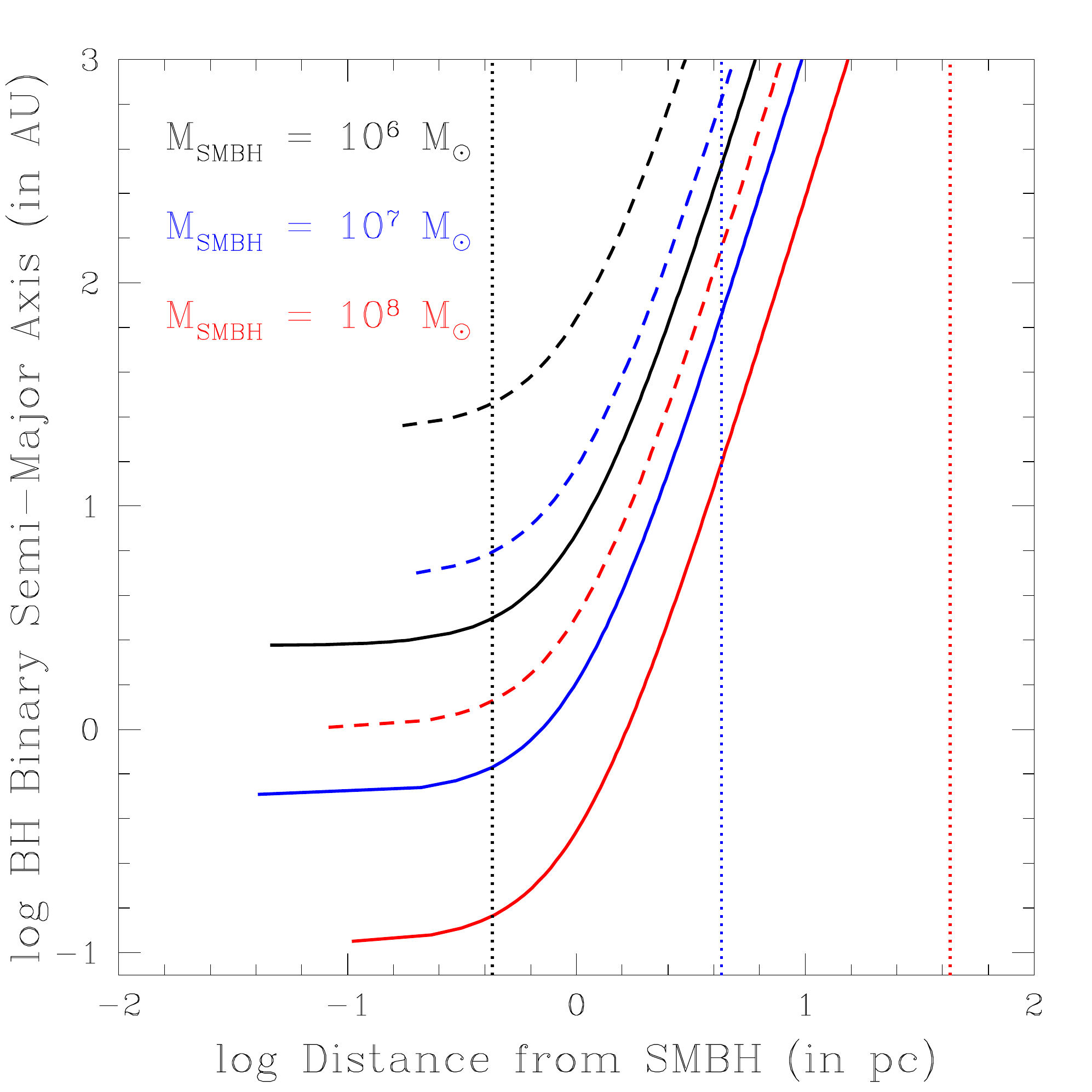}
\caption{The critical IMBH-SBH binary semi-major axis as a function of distance from the SMBH at which the mass segregation time-scale is roughly equal to the characteristic Lidov-Kozai time-scale.  The solid lines show the results assuming $M_{\rm IMBH} = 10^2$ M$_{\odot}$ and $M_{\rm BH} = 10$ M$_{\odot}$, and the dashed lines show $M_{\rm IMBH} = 10^3$ M$_{\odot}$ and $M_{\rm BH} = 10$ M$_{\odot}$.  The dotted vertical lines show the influence radii for different SMBH masses (10$^6$ M$_{\odot}$, 10$^7$ M$_{\odot}$ and 10$^8$ M$_{\odot}$).  Finally, we assume an eccentricty of 0.3 for the outer orbit of the SMBH-IMBH-SBH triplet.}
\label{fig:rcrit}
\end{figure}

This last mechanism for BH triplet formation occurs in situ and requires the prior presence of stellar-mass BHs in addition to a gaseous AGN disk, whereas this is not the case for the first two mechanisms. Here, a binary BH is (presumably) delivered to the outskirts of the central nuclear star cluster, before segregating inward due to two-body relaxation.  Hence, for the first two mechanisms, we naively expect any IMBH-BH binaries to have their orbital planes aligned isotropically relative to the central SMBH.  
To better quantify the competition between KL oscillations and mass segregation, we refer the reader to Figure~\ref{fig:rcrit}.  This shows the critical IMBH-BH binary semi-major axis as a function of distance from the central SMBH at which the characteristic time-scale for mass segregation is roughly equal to that for Kozai-Lidov oscillations. The former time-scale is given by:
\begin{equation}
\tau_{relax}(r) = 1.7 \times 10^5\ \mathrm{yr}\ \Big( \frac{\bar{m}}{m_{\rm b}} \Big) \Big( \frac{r}{1\ \mathrm{pc}} \Big)^{3/2} N(r)^{1/2} \Big( \frac{\bar{m}}{\mathrm{M}_{\odot}} \Big) \,
\label{eqn:trelax}
\end{equation}
where $\bar{m}$ is the average stellar mass in the cluster, $m_{\rm b}=M_{IMBH}+M_{SBH}$ is the mass of the IMBH-BH binary, $r$ is the distance of the IMBH-BH binary centre of mass from the central SMBH and $N(r)$ is the number of stars within a distance $r$ from the central SMBH.  

Now, the characteristic time-scale (at the approximation of the quadrupole level) for eccentricity oscillations due to Kozai-Lidov cycles is \citep{koz62,lid62}:
\begin{equation}
\tau_{LK}(r) = P_{\rm in}\Big( \frac{m_{\rm b}}{M_{\rm SMBH}} \Big)\Big( \frac{r}{a_{in}} \Big)^3(1-e_{out}^2)^{3/2},
\label{eqn:tauLK}
\end{equation}
where $m_{\rm b}$ is the mass of the IMBH-BH binary, $M_{\rm SMBH}$ is the mass of the central SMBH, $P_{in}$ and $a_{in}$ are the orbital period and semi-major axis, respectively, of the IMBH-BH binary and $e_{out}$ is the orbital eccentricity of the IMBH-BH binary centre-of-mass orbit about the SMBH.  

Setting Eq.~\ref{eqn:trelax} equal to Eq.~\ref{eqn:tauLK}, we can solve for the critical IMBH-BH binary semi-major axis at which these two time-scales are roughly equal as a function of the distance of the IMBH-BH binary centre of mass from the SMBH. Assuming a Plummer density profile for the central nuclear cluster with a scale radius of 1 pc and a central mass density of $\rho_{\rm 0} = 10^7$ M$_{\odot}$ pc$^{-3}$, we first substitute $N(r) = 4/3{\pi}r^3\rho_{\rm 0}/\bar{m}$ in to Equation~\ref{eqn:trelax}.  

The results of this exercise are shown in Figure~\ref{fig:rcrit}. The solid and dashed lines correspond to cases where the two aforementioned time-scales are equal, whereas the dotted vertical lines correspond to the influence radius $r_{\rm inf} = Gm_{\rm SMBH}/\sigma^2$, where $\sigma$ is the velocity dispersion of the surrounding nuclear star cluster, and we assume $\sigma = 100$ km s$^{-1}$. Beyond the influence radius, Kozai-Lidov oscillations should not operate, since the gravitational potential of the central SMBH does not dominate over the local nuclear cluster potential.  

The take-away message from Figure~\ref{fig:rcrit} is that \textit{wider more massive} IMBH-SBH binaries should segregate deeper in the host nuclear star cluster potential before becoming active in the Kozai-Lidov regime (i.e., the time-scale for KL oscillations becomes short compared to the local mass segregation time-scale).  

\section{Numerical method}
\label{sect:method}

\begin{figure} 
\centering
\includegraphics[scale=0.5]{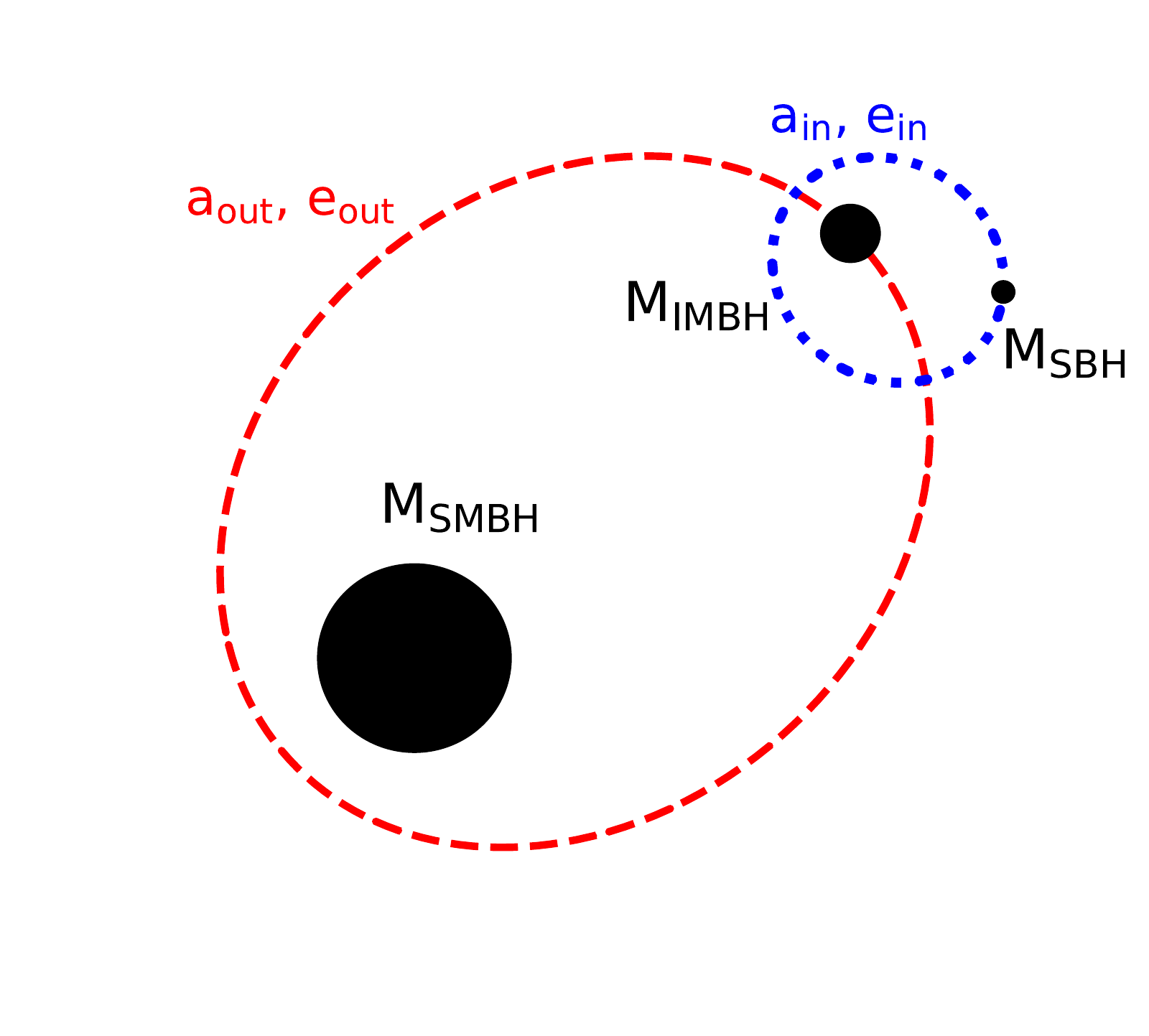}
\caption{The three-body system studied here. We indicate the mass of the SMBH as $M_\mathrm{SMBH}$, the mass of the IMBH as $M_\mathrm{IMBH}$ and the mass of the SBH as $M_\mathrm{SBH}$. The semimajor axis and eccentricity of the outer orbit are $a_{out}$ and $e_{out}$, respectively, while for the inner orbit they are $a_{in}$ and $e_{in}$, respectively.}
\label{fig:threebody}
\end{figure}

We study the fate of IMBH-SBH binaries in galactic nuclei that host a massive black hole, as a function of the the IMBH orbit and SMBH-IMBH mass ratio. As shown in Fig. \ref{fig:threebody}, we study a three-body system comprising of the inner binary IMBH-SBH, and an outer binary comprised of the SMBH and the centre of mass of the IMBH-BH binary. We name the mass of the SMBH as $M_\mathrm{SMBH}$, the mass of the IMBH as $M_\mathrm{IMBH}$ and the mass of the SBH as $M_\mathrm{SBH}$. The semimajor axis and eccentricity of the inner orbit are $a_{in}$ and $e_{in}$, respectively, while for the outer orbit these are $a_{out}$ and $e_{out}$, respectively.

\begin{table*}
\caption{Models: name, mass of the SMBH ($M_\mathrm{SMBH}$), mass of the IMBH ($M_\mathrm{IMBH}$), semimajor axis of the outer orbit ($a_{out}$), eccentricity of the outer orbit ($e_{out}$).}
\centering
\begin{tabular}{lcccc}
\hline
Name &	$M_\mathrm{SMBH}$	(M$_\odot$) & $M_\mathrm{IMBH}$	(M$_\odot$) & $a_{out}$	(pc) & $e_{out}$\\
\hline\hline
MW	& $4\times 10^6$ & $5$-$10\times 10^3$ & $0.1$ & $0.4$ \\
MW	& $4\times 10^6$ & $5\times 10^3$ & $0.05$-$0.1$-$0.5$ & $0.4$ \\
MW	& $4\times 10^6$ & $5\times 10^3$ & $0.1$ & $0$-$0.4$-$0.7$ \\
GN	& $1\times 10^8$ & $5$-$10\times 10^3$ & $0.1$ & $0.4$ \\
GN	& $1\times 10^8$ & $5\times 10^3$ & $0.05$-$0.1$-$0.5$ & $0.4$ \\
GN	& $1\times 10^8$ & $5\times 10^3$ & $0.1$ & $0$-$0.4$-$0.7$ \\
\hline
\end{tabular}
\label{tab:models}
\end{table*}

As discussed in Sect. \ref{sect:imbhnuclei}, if the relative inclination between the inner and outer orbits is in the active Kozai-Lidov regime, namely with an inclination angle (between the two orbital planes) in the window $40^\circ\lesssim i\lesssim 140^\circ$, the eccentricity and inclination of the inner orbit can experience periodic oscillations on a secular Kozai-Lidov time-scale (Eq.~\ref{eqn:tauLK}). We note that the exact size of the Kozai-Lidov angle window depends also on the physical parameters of the three objects, thus varying from case to case \citep{grish17,grish18}. On this typical time-scale, the relative inclination of the inner orbit and outer orbit slowly increases while the orbital eccentricity of the inner orbit decreases, and vice versa, conserving angular momentum \citep{nao16}. The eccentricity of the two orbits can be excited up to a maximum eccentricity determined by the initial inclination $i_0$
\begin{equation}
e_{in,max}=\sqrt{1-\frac{5}{3}\cos i_0^2}\ .
\label{eqn:emax}
\end{equation}
However, Kozai-Lidov cycles can be suppressed by additional sources of apsidal precession, such as relativistic precession or tidal bulges raised on the surfaces of the objects \citep*{nao16}. In the case of a SMBH-IMBH-SBH triple, the most significant mechanism to consider is general relativistic precession, which influences the dynamics on a typical time-scale
\begin{equation}
\tau_{GR}(r)=\frac{a_{in}^{5/2}c^2(1-e_{in}^2)}{3G^{3/2}m_{\rm b}^{3/2}}\ .
\end{equation}
In the region of the parameter space where $\tau_{KL}>\tau_{GR}$, the Kozai-Lidov oscillations of the orbital elements are damped by relativistic effects. 

Usually, calculations adopt the secular approximation to study hierarchical triples. Although faster and less computationally expensive, secular theory may fail to predict the maximum eccentricity excited by Kozai-Lidov oscillations suppressed by general relativistic effects.  The correct time evolution of such triples experiencing strong relativistic precession can only be accurately captured using sophisticated $N$-body integrators with regularization applied, in particular in critical cases such as when the eccentricity of the inner orbit becomes very large \citep{antognini14,anm14}. For these cases, direct $N$-body simulations, including Post-Newtonian (PN) terms, are required to follow accurately the orbits of the objects up to the final merger. 

We note that the exact distribution of inner and outer semi-major axis is quite unconstrained for an SMBH-IMBH-SBH triple system. Hence, we consider a quite wide range of initial conditions over which we average to derive the IMRI rate. The initial conditions for our $N$-body simulations have been set as follows (see also Table\,\ref{tab:models}):
\begin{itemize}
\item the mass of the SMBH is set to $M_\mathrm{SMBH}=4\times 10^6\msun$ (i.e., a Milky-Way like nucleus; Model MW) or $M_\mathrm{SMBH}=10^8\msun$ (i.e., a more massive galactic nucleus such as in M31; Model GN);
\item the mass of the IMBH is $M_\mathrm{IMBH}=5\times 10^3\msun$-$10^4\msun$;
\item the mass of the SBH is fixed to $M_\mathrm{SBH}=10\msun$;
\item the semimajor axis of the SMBH-IMBH orbit is in the range $\aout=0.05\pc$-$0.5\pc$;
\item the eccentricity of the SMBH-IMBH orbit is in the range $e_\mathrm{out}=0$-$0.7$;
\item the semimajor axis of the inner orbit is sampled uniformly within the Hill sphere of the IMBH at its orbital pericentre with respect to the SMBH
\begin{equation}
R_{H}=a_{out}(1-e_{out})\left(\frac{M_{IMBH}}{M_{SMBH}}\right)^{1/3}
\label{eqn:hills}
\end{equation}
\item the eccentricity of the inner orbit is sampled from a uniform distribution;
\item the initial mutual inclination $i_0$ between the inner and outer orbits is drawn from an isotropic distribution;
\item the initial phases $\Psi$ and $\Phi$ of the inner and outer orbits, respectively, are drawn randomly.
\end{itemize}

Given the above set of initial parameters, we integrate the triple SMBH-IMBH-SBH differential equations of motion
\begin{equation}
{\ddot{\textbf{r}}}_i=-G\sum\limits_{j\ne i}\frac{m_j(\textbf{r}_i-\textbf{r}_j)}{\left|\textbf{r}_i-\textbf{r}_j\right|^3}\ ,
\end{equation}
with $i=1$,$2$,$3$. The integrations are performed using the fully regularised \textsc{archain} code \citep{mik06,mik08,hellst10}. We include PN corrections up to order PN2.5. 

\section{Intermediate-mass ratio inspirals in galactic nuclei}
\label{sect:imris}

For each set of parameters, we run $1500$ simulations up to a maximum time $T=1$ Myr. This time-scale is roughly the Kozai-Lidov time-scale at the octupole level of approximation \citep{lin15,fraglei18}. In our simulations the IMBH-SBH binary has three possible fates: (i) the IMBH-SBH can be broken up by differential forces exerted by the SMBH, such that the SBH will either be captured by the SMBH or ejected from the galactic nucleus; (ii) the IMBH-SBH binary can survive on an orbit perturbed with respect to the original one; (iii) the IMBH-SBH pair merges producing an IMRI event. We distinguish among these possible outcomes by computing the mechanical energy of the SBH with respect to the IMBH at every integration time-step. If the relative energy becomes positive, we consider the IMBH-SBH unbound (case (i)), while if it remains negative we consider the IMBH-SBH survived (case (ii)). Finally, if the IMBH-SBH binary merges, which occurs if the Schawrzschild radii of the two BHs overlap directly, we have an IMRI (case (iii)).

\begin{figure} 
\centering
\includegraphics[scale=0.5]{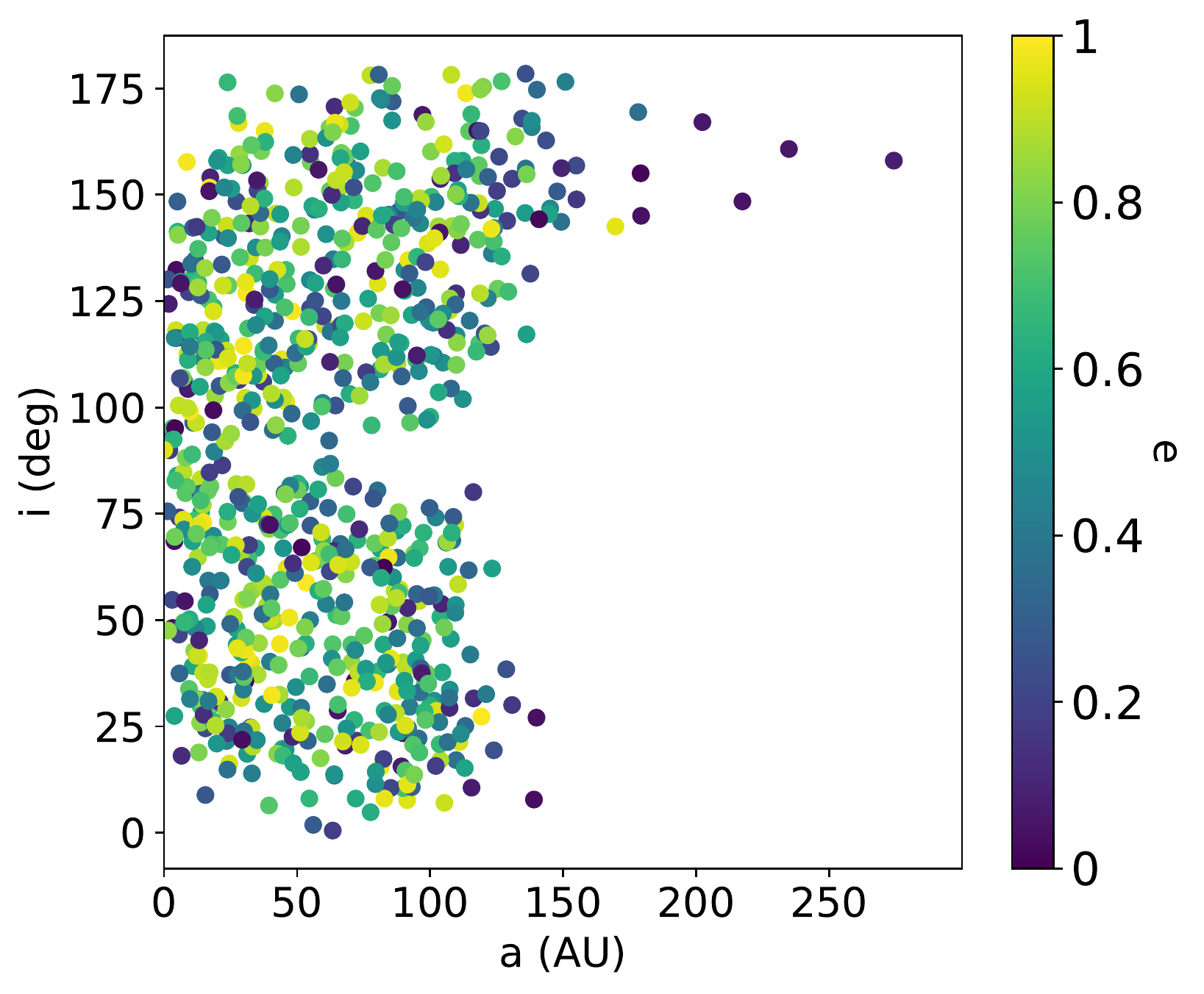}
\caption{Final distribution of surviving IMBH-BH binaries, when the mass of the SMBH is $M_\mathrm{SMBH}=1\times 10^8\msun$ and the mass of the IMBH is $M_\mathrm{IMBH}=5\times 10^3\msun$. The outer orbital semi-major axis and eccentricity are $a_{out}=0.1\pc$ and $e_{out}=0.4$, respectively.}
\label{fig:finalbhs}
\end{figure}

We illustrate in Fig.~\ref{fig:finalbhs} the final distribution of surviving IMBH-BH binaries, when the mass of the SMBH is $M_\mathrm{SMBH}=1\times 10^8\msun$ and the mass of the IMBH is $M_\mathrm{IMBH}=5\times 10^3\msun$. The outer orbital semi-major axis and eccentricity are initially $a_{out}=0.1\pc$ and $e_{out}=0.4$, respectively. Above $\sim 300$ AU, there are no stable orbits around the IMBH; IMBH-SBH binaries are broken up by the continuous differential gravitational pull of the SMBH. While the initial inner and outer orbit relative inclinations are distributed isotropically, the final distribution of surviving systems lacks highly inclined IMBH-SBH binaries \citep{grish17}. The lack of highly inclined systems shows the importance of the Kozai-Lidov mechanism \citep{fraglei18}. IMBH-SBH binaries that successfully undergo an IMRI event originally orbit in a plane highly inclined with respect to the outer orbital plane. In these binaries, the Kozai-Lidov mechanism governs the dynamics of the system and induces oscillations both in eccentricity and inclination, whenever not suppressed by relativistic precession. The majority of IMBH-SBH binaries that remain bound have typical semimajor axes $\lesssim 150$ AU and moderate orbital inclinations out of the active Kozai-Lidov window.

\begin{figure} 
\centering
\includegraphics[scale=0.55]{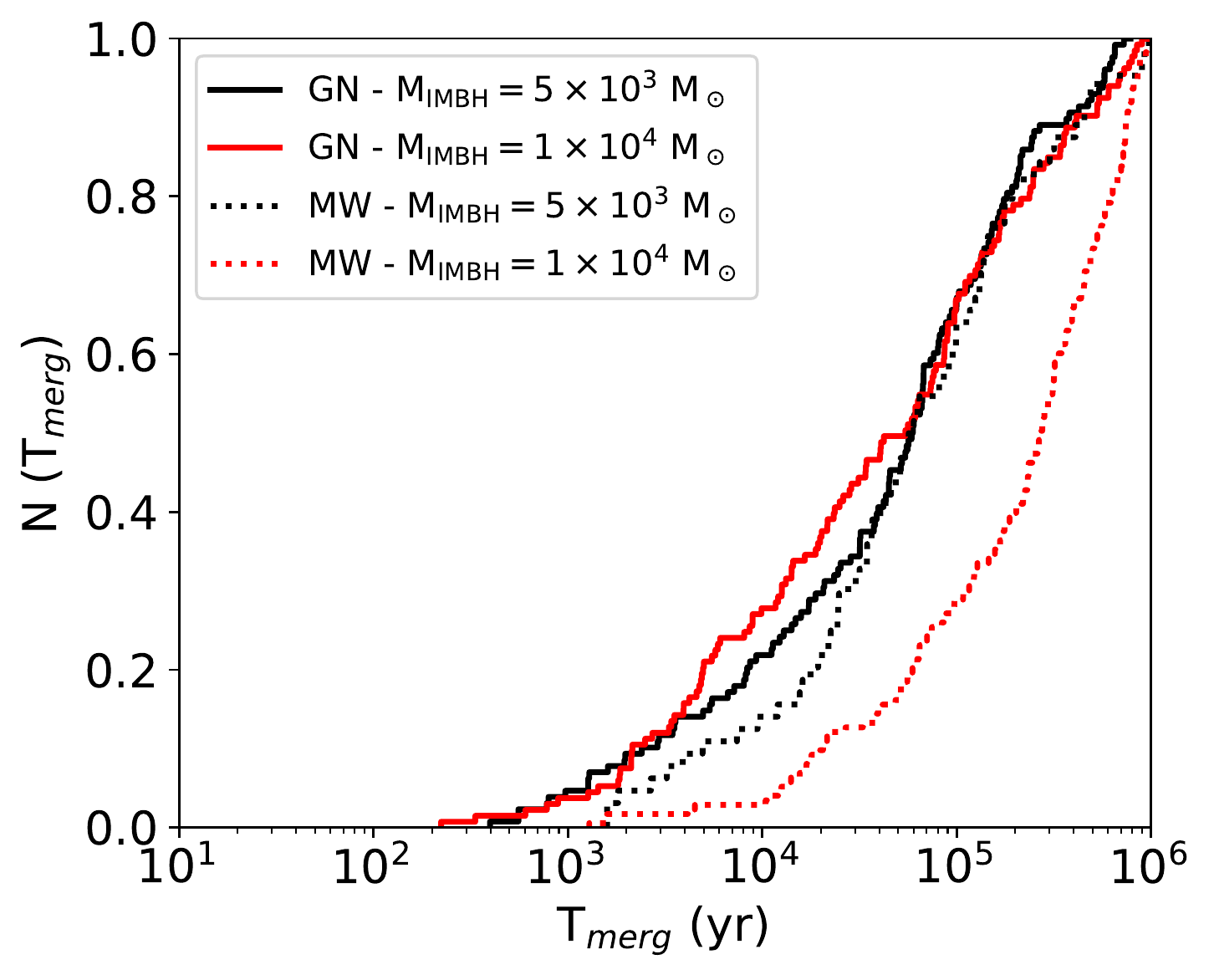}
\includegraphics[scale=0.55]{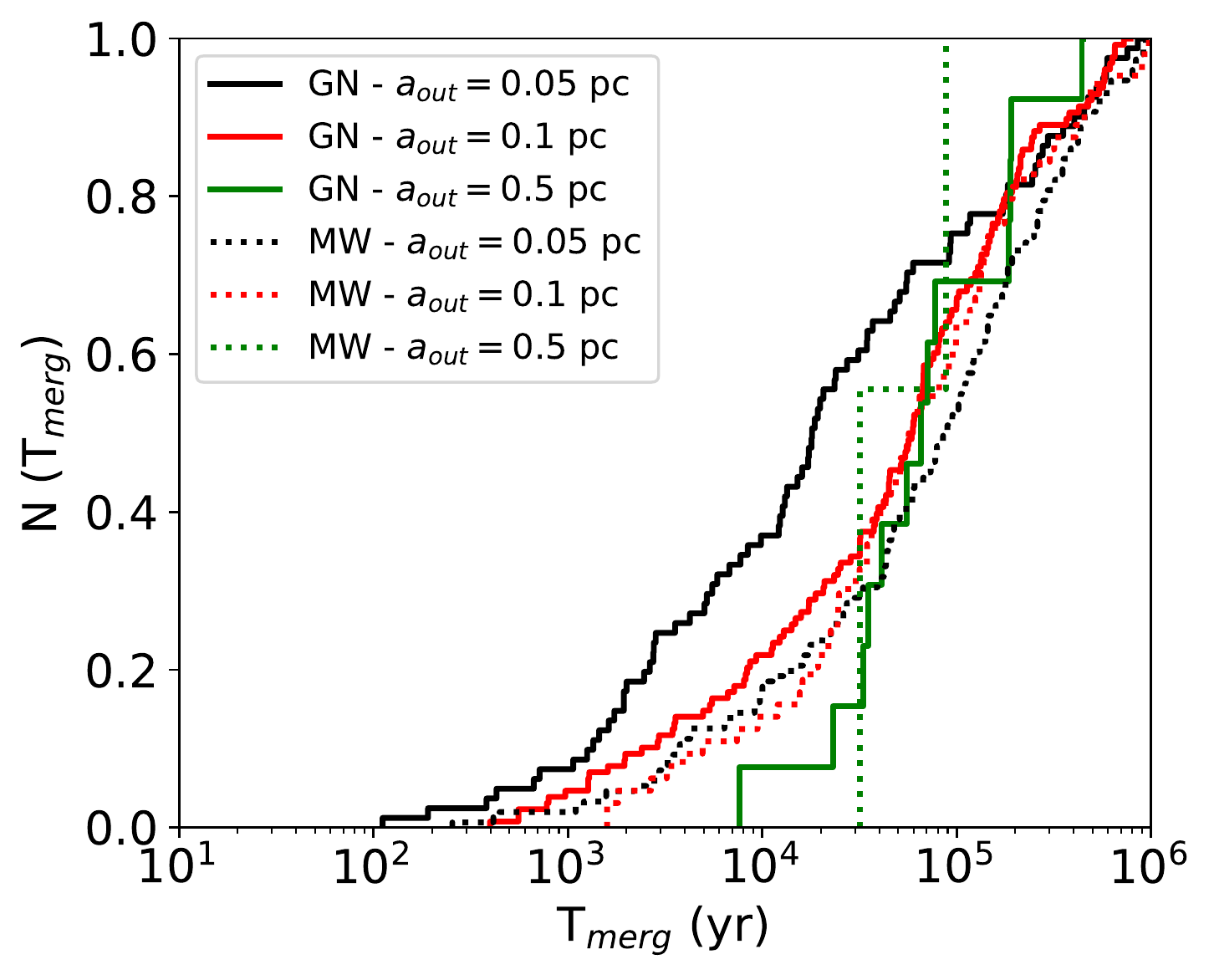}
\includegraphics[scale=0.55]{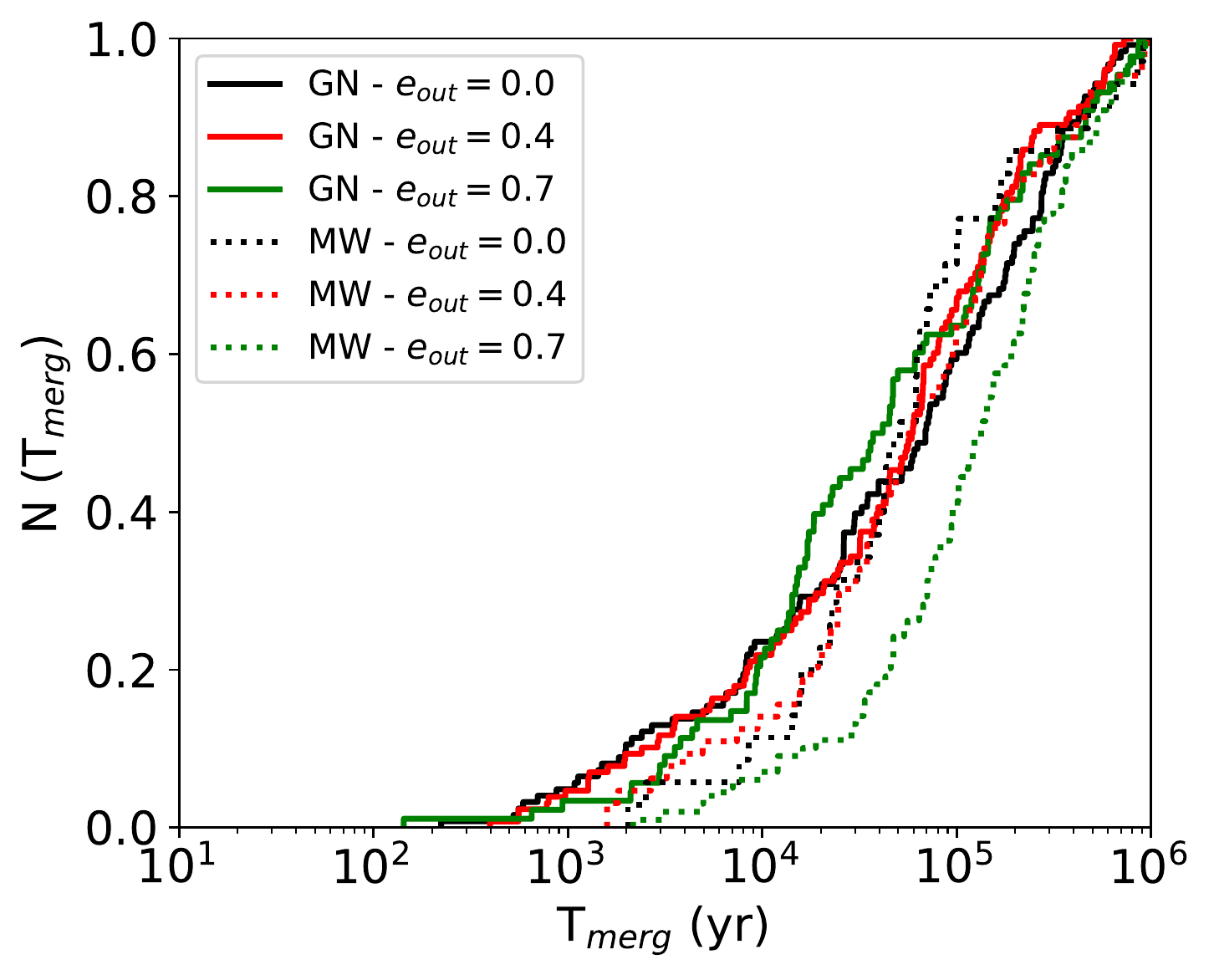}
\caption{IMRI merger time (T$_{merg}$) cumulative distributions for Models GN and MW. Top panel: $M_\mathrm{IMBH}=5\times 10^3\msun$-$1\times 10^4\msun$, $a_{out}=0.1\pc$ and $e_{out}=0.4$. Central panel: $M_\mathrm{IMBH}=5\times 10^3\msun$, $a_{out}=0.05\pc$-$0.1\pc$-$0.5\pc$ and $e_{out}=0.4$. Bottom panel: $M_\mathrm{IMBH}=5\times 10^3\msun$, $a_{out}=0.1\pc$ and $e_{out}=0.0$-$0.4$-$0.7$.}
\label{fig:tmerg}
\end{figure}

\subsection{Distributions of IMBH-SBH merger times}

\begin{figure*} 
\centering
\begin{minipage}{20.5cm}
\includegraphics[scale=0.55]{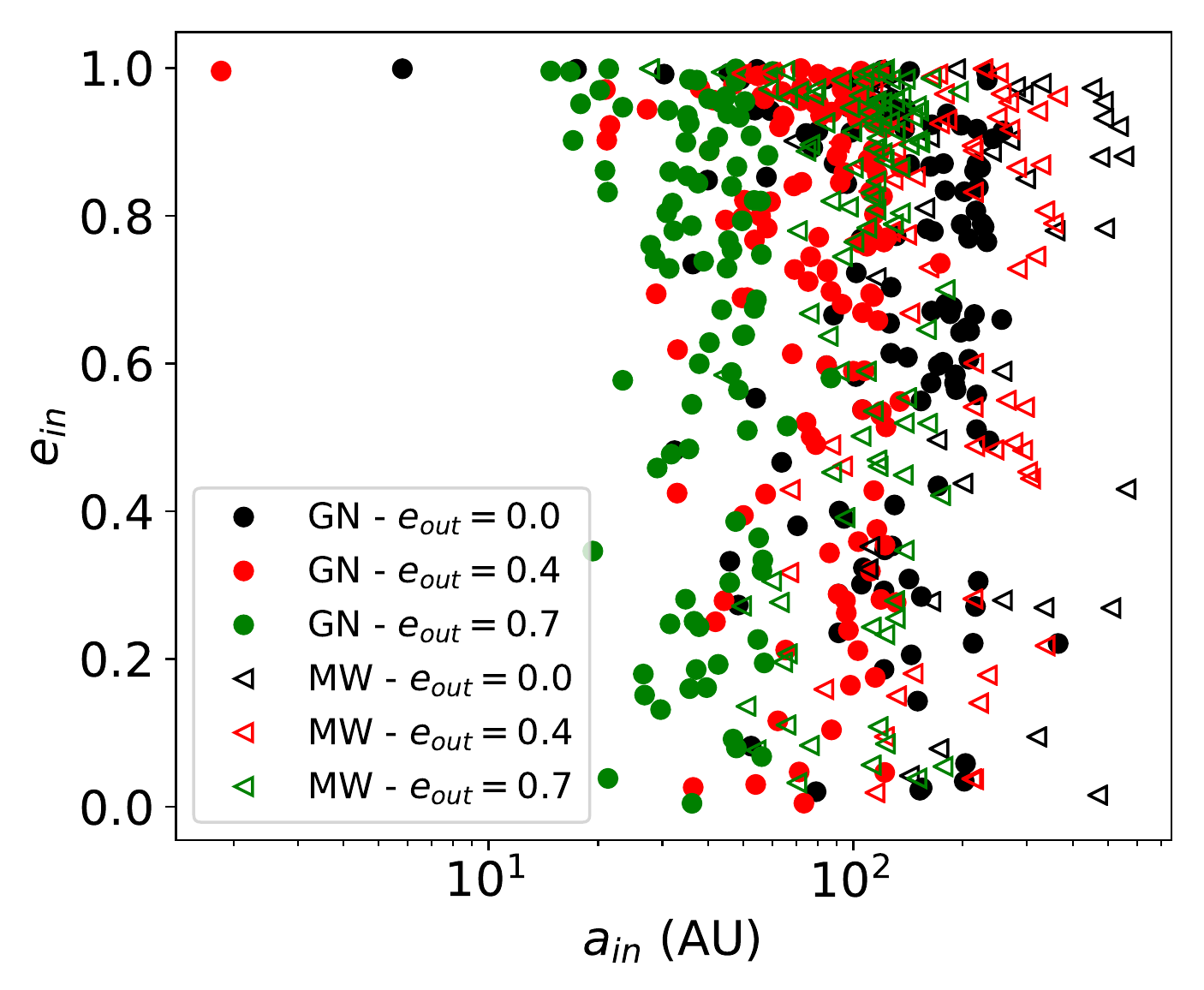}
\hspace{1.5cm}
\includegraphics[scale=0.55]{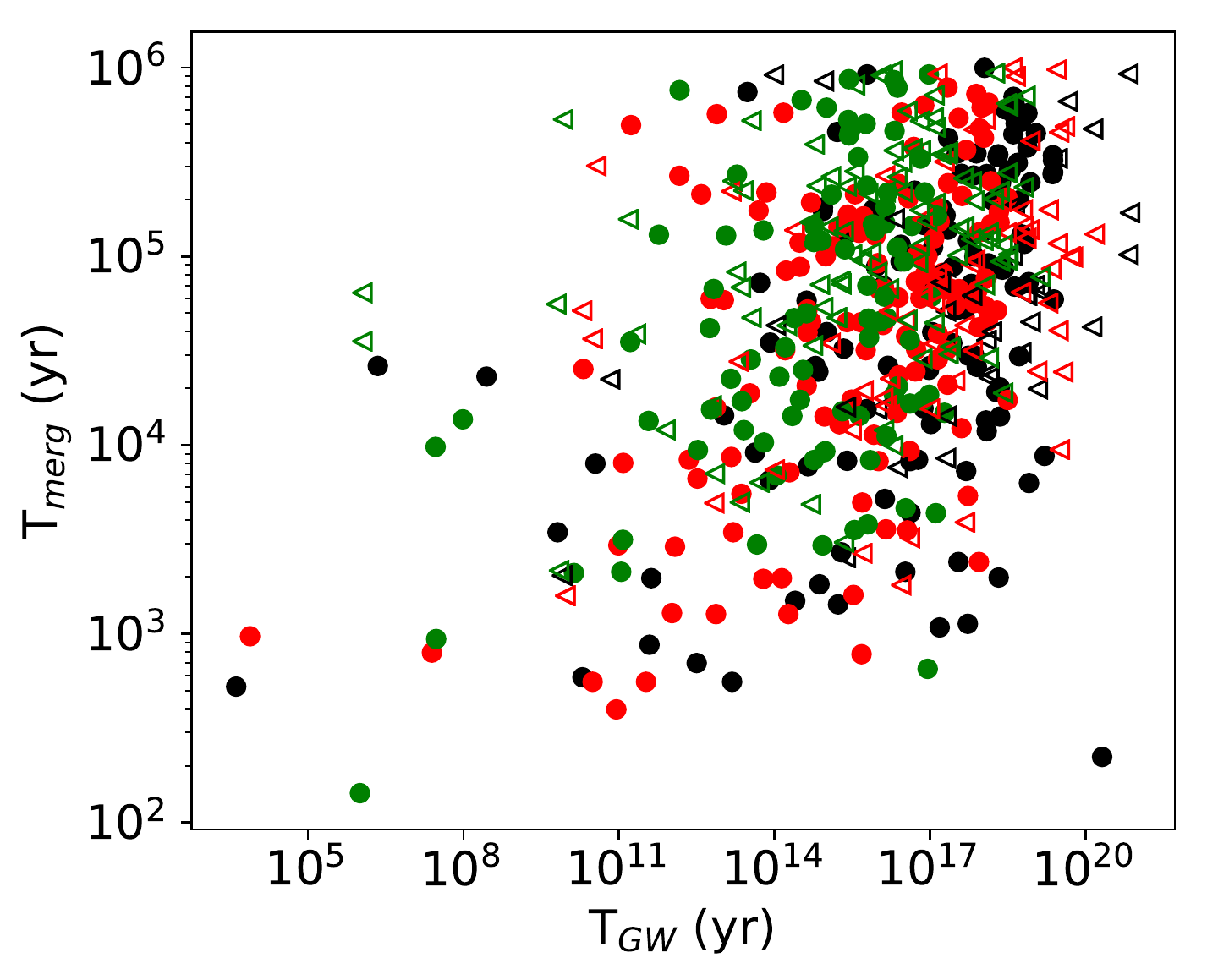}
\end{minipage}{20.5cm}
\caption{Left: Initial semi-major axes and eccentricities of the inner orbit that end up as IMRIs when $M_\mathrm{IMBH}=5\times 10^3\msun$ and the outer orbital semi-major axis and eccentricity are $a_{out}=0.1\pc$ and $e_{out}=0.0$-$0.4$-$0.7$, respectively. Right: merger time T$_{merg}$ as a function of the nominal \citet{pet64} GW merger time-scale T$_{GW}$, for the same IMBH-SBH binaries as shown in the left panel.}
\label{fig:tgwtmer}
\end{figure*}

In Fig.~\ref{fig:tmerg}, we show the cumulative distributions of IMRI merger times (T$_{merg}$) for Models GN and MW. In the top panel, we show the cumulative distribution for $M_\mathrm{IMBH}=5\times 10^3\msun$-$1\times 10^4\msun$, $a_{out}=0.1\pc$ and $e_{out}=0.4$. In Model GN, we do not find a significant difference for different IMBH masses.  This is likely due to the very small mass ratio $M_\mathrm{IMBH}/M_\mathrm{SMBH}=5$-$10\times 10^{-5}\msun$. In a Milky-Way like nucleus, this indicates a non-negligible shift of roughly one order of magnitude between the case $M_\mathrm{IMBH}=5\times 10^3\msun$ and $M_\mathrm{IMBH}=1\times 10^4\msun$. Since the mass ratio is larger than in the previous case, less-massive IMBH-SBH binaries are more significantly perturbed by the SMBH, thus merging on shorter time-scales. In the central panel, we plot the aforementioned distributions for the case $M_\mathrm{IMBH}=5\times 10^3\msun$, $a_{out}=0.05\pc$-$0.1\pc$-$0.5\pc$ and $e_{out}=0.4$. In both Model GN and Model MW, larger outer semi-major axes usually imply longer merger times, scaling the typical Kozai-Lidov time-scale as $\propto a_{out}^3$. Finally, in the bottom panel, we show the cumulative distributions for $M_\mathrm{IMBH}=5\times 10^3\msun$, $a_{out}=0.1\pc$ and $e_{out}=0.0$-$0.4$-$0.7$. In all models, we do not see significant differences in the merger time distributions. Larger outer orbital eccentricities imply a smaller apocentre distance for disruption due to it exceeding the IMBH Hill radius, thus the typical IMBH-SBH inner semi-major axis must be smaller. This implies both a longer Kozai-Lidov time-scale ($\propto a_{in}^{-3/2}$) and a smaller relativistic precession time-scale ($\propto a_{in}^{5/2}$), which may negate Kozai-Lidov resonances.

To help illustrate how perturbations from the SMBH reduce the merger time, we show in Fig.~\ref{fig:tgwtmer} the initial semi-major axes and eccentricities of the inner orbits that end up as IMRIs when $M_\mathrm{IMBH}=5\times 10^3\msun$ assuming that the initial outer orbital semi-major axis and eccentricity are $a_{out}=0.1\pc$ and $e_{out}=0.0$-$0.4$-$0.7$, respectively (left panel).  We also show the merger time T$_{merge}$ as a function of the nominal \citet{pet64} GW merger time-scale
\begin{equation}
T_{\mathrm{GW}}=\frac{3}{85}\frac{a^4 c^5}{G^3 M_{\mathrm{IMBH}} M_{\mathrm{SBH}} M}(1-e^2)^{7/2}\ ,
\label{eqn:peters}
\end{equation}
for the same IMBH-SBH binaries (right panel). Most of the IMBH-SBH binaries that successfully end up as IMRIs have typical semi-major axes $10\au\lesssim a_{in}\lesssim 300\au$ and $50\au\lesssim a_{in}\lesssim 700\au$ for $M_\mathrm{SMBH}=1\times 10^8\msun$ and $M_\mathrm{SMBH}=4\times 10^6\msun$, respectively. As discussed previously, most of these binaries have initial orbital planes highly inclined with respect to the outer orbital plane, thus being affected by Kozai-Lidov oscillations. Due to oscillations in the orbital elements, the IMBH-SBH binaries merge much faster than predicted by Eq.~\ref{eqn:peters}.

\subsection{IMRI rates in the local Universe}
Although we have explored a limited number of SMBH masses, we emphasize that the range is relatively representative of what is expected for galactic nuclei. Hence, in this section, we derive upper and lower limits for the IMRI rate in the local Universe, from which we infer the dependence of the rate on the distribution of SMBH masses in the nearby Universe, which remains poorly constrained.  

We calculate the IMRI rate as
\begin{equation}
\mathcal{R_\mathrm{IMRI}}=\xi n_{gal}\ \Gamma_{\mathrm{IMRI}}\ ,
\label{eqn:rategw}
\end{equation}
where $n_{gal}=0.02$ Mpc$^{-3}$ is the number density of galaxies \citep{cons15} and $\Gamma_{\mathrm{IMRI}}$ is the averaged merger rate of IMBH-SBH binaries from our simulations. In the previous equation, $\xi$ is the fraction of nuclei that host an SMBH-IMBH binary. In general, we find that larger IMBH masses and smaller IMBH-SBH semi-major axes lead to a larger IMRI rate. As well, smaller mass ratios $M_\mathrm{IMBH}/M_\mathrm{SMBH}$ imply a larger rate of such events. We find that for Model GN
\begin{equation}
\mathcal{R^{\mathrm{GN}}_\mathrm{IMRI}}=1.4\xi\ \mathrm{Gpc}^{-3}\ \mathrm{yr}^{-1}\ ,
\end{equation}
while for our Milky Way-like model
\begin{equation}
\mathcal{R^{\mathrm{MW}}_\mathrm{IMRI}}=0.74\xi\ \mathrm{Gpc}^{-3}\ \mathrm{yr}^{-1}\ .
\end{equation}
We note that a smaller SMBH implies a smaller rate ($\sim$ half) because of the smaller perturbations exerted by the SMBH on the IMBH-SBH binary.

From our rate computation, we find that the rate of merging IMBH-SBH binaries in galactic nuclei could be substantial. These IMRI events would be detectable in the near future by either \textit{LISA} or the Einstein Telescope out to a redshift $z\sim 1$-$2$ \citep{ama10,gai11}, thus offering the promise of confirming (or refuting) the existence of IMBHs in the local Universe.  If detected, these observations would potentially provide a large sample of IMBHs to study the formation, evolution, merger rates and scaling relations of IMBHs in extragalactic nuclei.

We note that in our study we take into account what is the fate of the IMBH-SBH systems when the IMBH-SBH has a long-term stationary orbit around the SMBH. As discussed in Sect. \ref{sect:imbhnuclei}, several processes can deliver the IMBH close to the centre of the galactic nucleus \citep{agu18,fragk18,secunda18}. The typical time-scale for the IMBH-SBH to reach a stable orbit ranges from a few Myrs to tens of Myrs. The previous estimates are sensitive to the initial conditions (e.g., the IMBH mass, the pericentre of the IMBH-SBH orbit, etc.). While we do not take into account the previous dynamical history that delivers the IMBH-SBH to such an orbit, we focus on its fate once this steady-state is reached. If we assume a typical time-scale of $\approx 10$ Myr to deliver an IMBH-SBH to the inner galactic nucleus once it has been delivered to the nucleus' outskirts (via GC infall, major/minor mergers, etc.), our rate would decrease by a factor of $\approx 10$. Moreover, over long time-scales the effect of the other stars surrounding the SMBH may play an important role, which deserves further consideration in future studies.

\section{Discussion and summary}
\label{sect:conc}

A distinctively unique signal for the presence of an IMBH could be GW radiation emitted during an IMRI event, when an SBH inspirals on to it \citep{fragk18}. Such events may happen within GCs, which may harbour IMBHs in their cores. If such GCs that harbour IMBH-SBH binaries are disrupted by galactic tides, any IMBH-SBH binaries will end up isolated in the field or in the host galaxy nucleus, if the host cluster orbit was such to deliver them there.  

In this paper, we study how IMBH-SBH binaries merge in galactic nuclei, driven by perturbations from the more massive SMBH. We consider different SMBH-IMBH mass ratios and orbital parameters, by means of high-precision direct $N$-body simulations, including a sophisticated regularization prescription and Post-Newtonian terms up to PN2.5 order. We show that the strong tidal field of the primary SMBH may lead to high variations in the eccentricities and inclinations of the IMBH-SBH binaries, which may result in IMRI events. We find that the rate of such extreme mass ratio inspirals could be as high as $\sim 1\ \mathrm{Gpc}^{-3}\ \mathrm{yr}^{-1}$ in the local Universe, and are expected to be associated with recent GC infall and/or major/minor galaxy mergers, making the observational signatures of such events (e.g., tidal debris) good diagnostics for searching for SMBH-IMBH-SBH mergers.

Some of these IMRI events could also be associated with an electromagnetic counterpart.  This could occur, for example, if the IMBH tidally captures (or even immediately disrupts) a main-sequence (MS) star, in analogy with \citet{lei14} for SBHs tidally capturing MS stars in GCs hosting a central IMBH. If tidal capture by the IMBH occurs, it will most likely happen close to the final IMBH-SBH inspiral. This is because the final orbital separation associated with the tidal capture event is typically factors of a few times the radius of the captured star \citep[e.g.][]{lee86}, and such a tidal capture will form a hierarchical triple composed of the IMBH-SBH inner binary with an outer MS companion (including the SMBH would make it a hierarchical quadruple system, technically) which requires a large ratio between the inner and outer orbital separations in order for the triple to be dynamically stable.  Kozai-Lidov oscillations of the outer SMBH-(MS-(IMBH-SBH)) triplet could then drive the MS star to very high eccentricities, leading to a tidal disruption event by either the IMBH or the SBH on time-scales of order years compared to when the IMBH-SBH inspiral occurs. Perturbations from background stars could also contribute to pumping the eccentricity on short time-scales, as found in \citet{lei14} using $N$-body simulations.

We calculate a tidal capture time-scale of order 10 Myr using Equation 5 in \citet{kalogera04}, and assuming a central stellar density of 10$^6$ pc$^{-3}$, a velocity dispersion of 100 km s$^{-1}$ (i.e., close to the influence radius where the local Keplerian velocity about the SMBH is comparable to the local velocity dispersion) and an IMBH mass of 10$^3$ M$_{\odot}$.  In the limit of a nearly inspiraled IMBH-SBH binary, GW emission dominates the rate of inspiral and we expect this phase to be much shorter than the tidal capture time-scale.  However, the mass segregation time-scale given by Equation~\ref{eqn:trelax} suggests a segregation time of order $\sim$ 1 Myr for a 10$^3$ M$_{\odot}$ IMBH in the MW Galactic Centre, corresponding to roughly a $\sim$ 10\% probability that the IMBH-SBH binary will tidally capture a MS triplet companion before reaching the active Kozai-lidov regime in orbit about the central SMBH.  Of course, the details of the efficiency of this scenario depend sensitively on the initial orbit of the IMBH-SBH binary, the total binary mass, the stellar density profile, etc.  Hence, we naively expect this phenomenon to be rare in the local Universe, but that it could reach a finite probability (since tidal capture/disruption is more likely than three-body scattering in the limit of a very compact IMBH-SBH binary orbit) if one assumes that nearly every galactic nucleus hosting an SMBH also has an IMBH-SBH companion \citep[e.g.][]{leigh16}.

With the above said, we strongly caution that there are many unknowns in performing more detailed calculations of this scenario, and some aspects of it may require significant fine-tuning to create the discussed configuration involving a quadruple MS star.  In particular, it is unlikely that the BHs can be much more massive than the MS stars they tidally capture, but the exact BH mass above which tidal capture can no longer occur, and tidal disruption will always occur instead, is highly uncertain \citep[e.g.][]{generozov18}.

\section{Acknowledgements}

GF acknowledges support from an Arskin postdoctoral fellowship and Lady Davis Fellowship Trust at the Hebrew University of Jerusalem. This research was partially supported by an ISF and an iCore grant. GF thanks Seppo Mikkola for helpful discussions on the use of the code \texttt{ARCHAIN}. Simulations were run on the \textit{Astric} cluster at the Hebrew University of Jerusalem.

\bibliographystyle{mn2e}
\bibliography{refs}

\end{document}